\documentclass[aps,prd,amssymb,nofootinbib]{revtex4}
\usepackage{graphicx,amsmath,subfigure}

\newcommand{\be}{\begin{equation}}
\newcommand{\ee}{\end{equation}}
\newcommand{\bel}[1]{\begin{equation}\label{#1}}
\newcommand{\ba}{\begin{eqnarray}}
\newcommand{\ea}{\end{eqnarray}}
\newcommand{\bal}[1]{\begin{eqnarray}\label{#1}}

\def\CQG{{\em Class.~Quantum Grav.~}}
\def\PRD{{\em Phys.~Rev.~D}}

\begin{document}

\title[Improved time-frequency analysis of EMRI signals in mock LISA data]{Improved time-frequency analysis of extreme-mass-ratio inspiral signals in mock LISA data}

\author{Jonathan R Gair}
\email{jgair@ast.cam.ac.uk}
\affiliation{Institute of Astronomy, Madingley Road, CB3 0HA, Cambridge, UK}

\author{Ilya Mandel}
\email{ilyamandel@chgk.info}
\affiliation{Department of Physics and Astronomy, Northwestern University, 2131 Tech Dr., Evanston, IL 60208}

\author{Linqing Wen}
\email{lwen@cyllene.uwa.edu.au}
\affiliation{Max-Planck-Institut f\"{u}r Gavitationsphysik, Am M\"{u}hlenberg 1, 14476, Potsdam, Germany}
\affiliation{Division of Physics, Mathematics, and Astronomy, Caltech, Pasadena, CA 91125, USA}
\affiliation{School of Physics, M013, The University of Western Australia, 35 Stirling Highway, Crawley, WA  6009, Australia}

\date{\today}

\begin{abstract}  
The planned Laser Interferometer Space Antenna (LISA) is expected to detect gravitational
wave signals from $\sim 100$ extreme-mass-ratio inspirals (EMRIs) of stellar-mass 
compact objects into massive black holes.   The long duration and large parameter space of EMRI signals makes data analysis for these signals a challenging problem.  One approach to EMRI data analysis is to use time-frequency methods. This consists of two steps: (i) searching for tracks from EMRI sources in a time-frequency spectrogram, and (ii) extracting parameter estimates from the tracks. In this paper we discuss the results of applying these techniques to the latest round of the Mock LISA Data Challenge, Round 1B. This analysis included three new techniques not used in previous analyses: (i) a new Chirp-based Algorithm for Track Search for track detection; (ii) estimation of the inclination of the source to the line of sight; (iii) a Metropolis-Hastings Monte Carlo over the parameter space in order to find the best fit to the tracks.
\end{abstract}

\maketitle

\section{Introduction}

The extreme-mass-ratio inspirals (EMRIs) of compact objects (white dwarfs, neutron stars or stellar-mass black holes) with
mass $m \sim 1 - 10\ M_\odot$ into massive black holes with mass $M \sim 10^6\ 
M_\odot$ can serve as excellent probes of strong-field general relativity (see \cite{Pau} and references therein for details). The detection of gravitational waves from EMRIs is one of the main targets for the planned gravitational-wave detector LISA (Laser Interferometer Space Antenna) \cite{LISA}. Astrophysical uncertainties mean that rate predictions are somewhat uncertain, but tens to thousands of EMRIs may be observed over the duration of the LISA mission \cite{Pau}.

Detection of EMRIs is not an easy task, however.  The expected gravitational wave (GW) signals from EMRIs will be buried in instrumental noise and a foreground of galactic white-dwarf binaries. The instantaneous signal amplitude will typically be an order of magnitude smaller than the amplitude of noise and foreground fluctuations in the detector at the same frequency. Matched filtering allows the identification of weak sources buried in noise, but this relies on the generation of templates of possible signals present in the data. The EMRI parameter space is fourteen dimensional. Following the notation of Barack and Cutler~\cite{AK}, these parameters are: the central black hole mass, $M$, the mass of the inspiraling object, $m$, the dimensionless spin parameter of the central black hole, $S$, the orbital frequency, $\nu_0$, and orbital eccentricity, $e_0$, specified at some moment of time, $t_0$, the inclination of the orbit to the equatorial plane of the central black hole, $\lambda$, the ecliptic latitude, $\beta$, and longitude, $\phi_S$, of the source on the sky, the orientation of the spin axis of the central black hole, $\theta_K$ and $\phi_K$, the distance to the source, $D$, and three phase angles, $\Phi_0$, $\gamma_0$ and $\alpha_0$, which specify, respectively, the azimuthal orbital phase, periapsis precession phase, and phase of the orbital-plane precession at $t=t_0$.  The large dimensionality of the EMRI parameter space, combined with the long observation timescale ($\sim$ year), makes fully coherent matched filtering using a template bank impractical as a detection method~\cite{gairetal}.

Time-frequency searches are one possible alternative to coherent matched filtering.  These have a much lower computational cost, albeit at the price of lower sensitivity.  Time-frequency methods consist of building a spectrogram of the signal by dividing it into shorter segments and performing a Fourier transform on each of these segments, identifying possible tracks in the spectrogram, and then finally estimating source parameters from the tracks. We previously used time-frequency searches to analyse the Mock LISA Data Challenge (MLDC)~\cite{Arnaud1} Round 2 data set~\cite{proc13}, and found that we were able to get good parameter recovery. That previous analysis used the Hierarchical Algorithm for Clusters and Ridges (HACR)~\cite{hacr07} to identify tracks by clustering bright pixels in binned spectrograms,  followed by a simple parameter estimation routine that used the track shape to infer the six intrinsic EMRI parameters $(M,m,S,\lambda,e_0,\nu_0)$ and used the power variation along the tracks to estimate the two sky position angles~\cite{proc13}.

The HACR algorithm looks for clusters of any shape, and uses a binning technique that is based on rectangular pixel bins~\cite{hacr07,wengair05,gairwen05}. This is a rather inefficient way to search for EMRI tracks, which have a characteristic chirping shape. The most recent release of the MLDC, Round 1B, was a repeat of the type of EMRI data previously released in Round 2~\cite{Arnaud2}. For this analysis we developed and used an improved track detection algorithm, which is tuned to look for EMRI type signals. We call this algorithm the Chirp-based Algorithm for Track Search (CATS; see \cite{CATS} for additional details).  For the Round 1B analysis, we also applied parameter-estimation methodology that had been improved relative to the techniques used in Round 2~\cite{proc13}. The improvements included an estimation of the inclination of the source to the line of sight and a Metropolis-Hastings Monte Carlo search to find the parameters that best reproduced the detected tracks.  The MLDC Round 1B data set consisted of five separate EMRI challenge data sets, 1B.3.1-1B.3.5; each challenge set comprised a single EMRI signal embedded in noise, with signal-to-noise ratio (SNR) between 40 and 110 \cite{Arnaud2}. We were able to detect EMRI signals in all of these data sets, and estimated 9 parameters for each challenge data set: the six intrinsic parameters --- $M,m,S,\lambda,e_0$ and $\nu_0$, the two sky direction angles, $\beta$ and $\phi_S$, and the angle $\kappa$ between the SMBH spin vector and the line of sight to the source.  We also independently estimated the plunge time, but this is determined by the other intrinsic parameters and was used to improve the determination of those parameters. Overall, the parameter recovery from the time-frequency techniques was very good and better than achieved in the Round 2 analysis~\cite{proc13}. 

The paper is organised as follows.  In Section~\ref{sec:search} we describe the new track detection techniques and in Section~\ref{sec:estimate} we describe the improvements to our parameter estimation algorithms. In Section~\ref{sec:results} we report the results of our analysis of the Round 1B data and compare these to the true parameters of the challenge data sets. Section~\ref{sec:future} includes a summary and a discussion of possible future improvements to the search pipeline.

\section{Track detection}\label{sec:search}
Our track search for the MLDC Round 2 analysis used HACR~\cite{hacr07,proc13}. We made a first cut at the Round 1B analysis using the same technique. We were able to detect several bright tracks for sources 1B.3.1, 1B.3.2 and 1B.3.3, and a cluster for 1B.3.4. The 1B.3.4 detection was very noisy, which was disappointing since this was a high SNR source. This failure prompted us to look at improved techniques more tuned to EMRI detection.

\subsection{Chirp Track Search}
HACR does not use all of the available information about the signal.  Specifically, we expect the tracks in the time-frequency spectrogram to be chirping curves characterized by three parameters: the frequency $f$ and its first two time derivatives $\dot{f}$ and $\ddot{f}$.  (The third time derivatives of frequency along the tracks are vanishingly small for low-eccentricity signals like those in challenges 1B.3.1-1B.3.3, and can be neglected for challenges 1B.3.4 and 1B.3.5 if sufficiently short sections of the tracks are considered.)  Moreover, the first and second time derivative of frequency must be positive.  This additional information can help to detect tracks.  In fact, when we scanned the time-frequency spectrogram by eye, we found some tracks that HACR didn't report, because we intuitively apply this additional information about the expected track shape when performing a visual search for tracks.

To make use of this information, we built a Chirp-based Algorithm for Track Search (CATS; see \cite{CATS} for additional details) to aid in track detection.  This algorithm works as follows:
\begin{itemize}
\item Construct $A$ and $E$ Time Delay Interferometry (TDI) channels for the challenge data sets.
\item For each channel, construct a spectrogram by dividing the data into time segments of equal duration, then Fourier transforming the data in each time segment after multiplying it by a Hanning window to reduce edge effects.  The time segment durations were $32 \times 10^5$ seconds for challenge 1B.3.1,  $8 \times 10^5$ seconds for challenges 1B.3.2 and 1B.3.3, and  $2 \times 10^5$ seconds for challenges 1B.3.4 and 1B.3.5.
\item Whiten the two spectrograms by dividing the signal power in each pixel by the expected LISA noise power spectral density, then construct a joint spectrogram by adding the two normalized spectrograms. 
\item Define the starting and ending time of a chirp search.  Construct a three-dimensional grid of parameters in the $(f, \dot{f}, \ddot{f})$ space.  For each point in the parameter space, build a potential track.  Then add up the power in all pixels along this track between the starting and ending time of the search to obtain the total track power.  (This is, effectively, a variation on a template-based matched-filtering search in the $(f, \dot{f}, \ddot{f})$ space.)
\item Find the brightest track and claim it as a detection.  Set the power in all pixels along this track in the spectrogram to large negative values in order to keep future detected tracks from intersecting the current track.  Repeat this step until all tracks are detected.
\end{itemize}

For this round, we did not attempt to search the whole spectrogram using this technique, nor did we search over the entire parameter space of $(f, \dot{f}, \ddot{f})$. Instead we used the HACR algorithm to identify clusters in the spectrogram arising from the presence of signals in the data stream, and the approximate shapes of these tracks. We then targeted those areas for follow-up with the CATS algorithm. At the time of the Round 1B analysis, we had not robustly determined the correct thresholds for track detection.  In the future, the track power threshold will be set based on a desired value of the false alarm probability. For this round, we set thresholds by hand, but performed a number of ``sanity checks'' to decide whether a given track detection was to be believed.  These checks included:
\begin{itemize}
\item Compare the detected tracks with a visual observation of the spectrogram.
\item Check if the tracks have sidetracks due to FFT bleeding, i.e., tracks that are parallel to them but one pixel off on each side.
\item Check if there is a clear harmonic structure typical to EMRIs, with sidebands caused by orbital-plane precession, i.e., several tracks that are nearly parallel (same $\dot{f}$) with a fixed offset from each other.
\item Check if the track parameters are reasonable: e.g., if presumed sections of a track are found at early and late times, the frequency and frequency derivative at the later time should be greater than those at the earlier time.
\end{itemize}

In all the Round 1B challenge data sets, CATS successfully found all tracks that were detected by HACR or could be made out by eye.  For some of the challenges, CATS found additional sidebands and CATS was also able to find several tracks in the Round 1B.3.5 data set, in which HACR was unable to identify any interesting clusters. This was particularly pleasing, since previous work~\cite{wengair05,gairwen05,hacr07} had indicated that sources of low SNR, like 1B.3.5, would be out of reach of time-frequency methods.

We estimated the systematic uncertainties of CATS by comparing the results of track searches using different time steps and different gridding of the chirp parameter space.  These comparisons suggest uncertainties on the order of $\Delta f \sim 10^{-7}$ Hz, $\Delta \dot{f} \sim 10^{-14}$ Hz/s, and $\Delta \ddot{f} \sim 10^{-21}$ Hz/s$^2$.  Of course, the inherent errors in chirp parameter estimation of low-SNR signals, particularly for short tracks, could be much greater than these systematic uncertainties.  

\subsection{Radon Transform Search}
We have also been developing and testing another time-frequency technique that utilizes the modified Radon transform. For this search, we sum power in pixels along expected tracks in the time-frequency spectrogram of a given source.  The expected tracks are calculated by integrating the known evolution equations for EMRIs for a given set of initial parameters.  Tracks of the harmonics and sidebands of orbital frequencies are included in the summation, and we search through the whole parameter space for the maximum signal-to-noise ratio.  This technique was not fully developed at the time of submission, so although the results were compared to the results obtained from the CATS algorithm, only the latter results were submitted for this round. In the future, further improvement of the Radon transform algorithm will be to needed in order to overcome problems of getting stuck in local maxima of the parameter space.  We are also investigating similar techniques using the Hough transform which sums over number counts (instead of powers) of bright pixels along the track.   A search using the Radon or Hough transform normally takes minutes to complete on a laptop when a good initial guess has been made, or hours starting from a poor guess.

\section{Parameter Extraction}\label{sec:estimate}
Parameter extraction based on the tracks identified by CATS consisted of two steps.  In the first step we made rough estimates of the parameters. This was the approach used for parameter extraction in our analysis of the Round 2 data sets~\cite{proc13,expowsymp}. An EMRI waveform is characterized by three fundamental frequencies --- the orbital frequency, $\nu$, the perihelion precession frequency, $f_{\gamma}=\dot{\gamma}/2\pi$ and the frequency of precession of the orbital plane, $f_{\alpha}=\dot{\alpha}/2\pi$. Each track identified in the spectrogram is a harmonic of these three frequencies characterized by three integers, $(n,l,k)$, with $f = n \nu + l f_\gamma + k f_\alpha$. The analytic kludge (AK) waveforms of Barack and Cutler~\cite{AK}, which is the model used to generate the MLDC data sets, are quadrupole in nature and precession effects are included in an ad hoc way by precessing the observer about the source. This restricts $|l|, |k| \leq 2$ and, in practice, harmonics with $l \neq 2$ are suppressed relative to the $l=2$ harmonics and are not detected by a time-frequency analysis.

If three tracks are detected at a given time --- two $n$-harmonics plus a $k$ sideband for one of those harmonics, then the separation of the $n$ harmonics is a measure of $\nu$, the separation of the $k$ sideband is a measure of $f_{\alpha}$ and, if the harmonic numbers of one of the harmonics can be guessed, the frequency of that harmonic then gives $f_{\gamma}$. For a specification of $e_0$ and $\lambda$, $f_\alpha$ determines $S$, $f_\gamma$ determines $M$ and $\nu_0$ is a parameter that has been measured directly. The rate of change of frequency of one harmonic, $\dot{f}$, then determines $m$. If the plunge time and the frequency of at least one harmonic at some other time can also be estimated, it is possible to iterate on $e_0$ and $\lambda$ to determine all six intrinsic parameters for the source.

The motion of the LISA detector around the sun leads to a modulation in the detector response that depends on sky position. From the power variation along a track it is therefore possible to estimate the sky position of the source, $(\beta, \phi_S)$ by using the low-frequency approximation to the LISA response~\cite{cutler98, proc13}. For the Round 1B analysis, we also tried to use the relative power in the sidebands to estimate the orientation of the source relative to our line of sight. Physically, the relative sideband power can only depend on the inclination of the source to the line of sight, $\kappa$, which is given in terms of the standard waveform parameters by $\cos\kappa = \sin(\beta)\cos(\theta_K)+\cos(\beta)\sin\theta_K\cos(\phi_S-\phi_K)$. The relative harmonic power depends only on the absolute value of $\cos\kappa$, and so we can choose $\kappa \in [0,\pi/2]$. It is not possible to determine the two source orientation angles, $(\theta_K, \phi_K)$, from the single number $\kappa$ and so for the MLDC entry we returned $\kappa$ as an additional parameter.

\begin{figure}
\begin{center}
\includegraphics[keepaspectratio=true, height=\textwidth, angle=-90]{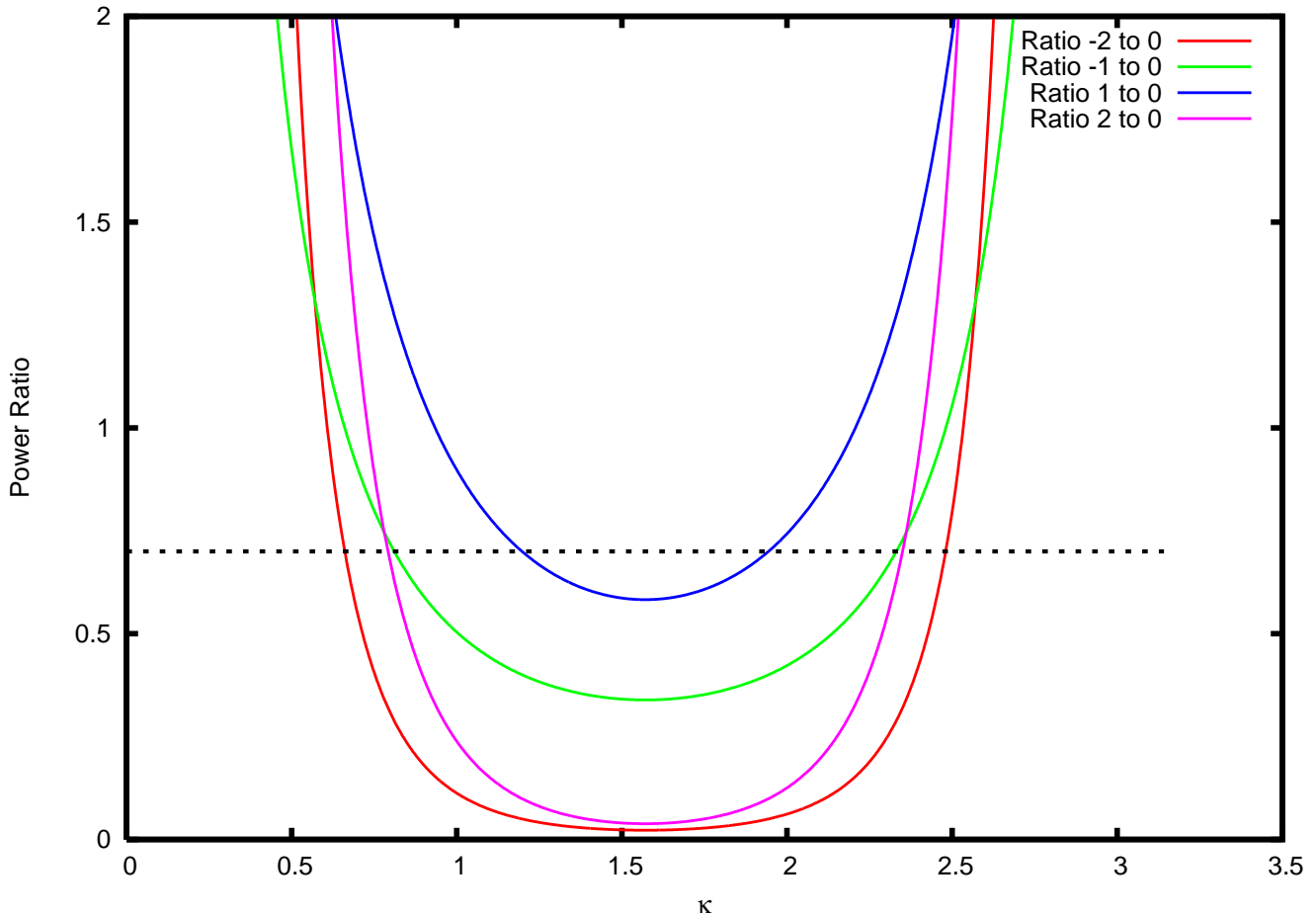}
\end{center}
\caption{Power in $k\neq0$ sidebands relative to power in $k=0$ sideband for challenge source 1B.3.2. The true parameters were $\lambda=1.44$, $\kappa=1.295$. Our estimated inclination was $\kappa=1.262$.}
\label{kapest}
\end{figure}

It turns out that, for a fixed value of $n$, the relative power in different $k$ harmonics depends only on $\kappa$ and the inclination of the orbit, $\lambda$ (more details will be given in~\cite{CATS}). The power ratio in sidebands can thus be used to estimate $\kappa$ and also to improve the determination of $\lambda$. This is illustrated in Figure~\ref{kapest}. The curves in this figure show the power in the various sidebands relative to the power in the $k=0$ harmonic as a function of $\kappa$, for a fixed value of $\lambda=1.44$. The horizontal dashed line shows the power ratio $k=1 : k=0$ estimated from the spectrogram, from which we infer an estimate $\kappa=1.262$. In the analysis of the Round 1B data, we found $\kappa\approx 1$ for all the data sets (specifically $\kappa = 1.15, 1.262, 0.99, 1.04, 1.1$ for Challenges 1B.3.1--1B.3.5 respectively). This suggested the determination of $\kappa$ was weak. The true values were $\kappa=0.443,1.295, 1.04, 1.17, 0.912$ respectively, which implies the measurement was quite accurate, except for 1B.3.1 in which the sideband detected was very weak, and so the harmonic power ratio was noise dominated. The value $\kappa \approx 1$ seems to be typical for points drawn from the MLDC priors, and the t-f spectrogram has limited ability to pin $\kappa$ down more accurately. However, the power ratio did help to improve the estimation of $\lambda$ for these data sets.

After determining rough parameter estimates via the above methods, the second stage of parameter estimation, which was used for the first time in this round of the MLDC, was to refine the estimates using an implementation of the Metropolis-Hastings Monte Carlo (MHMC) algorithm. The MHMC algorithm is described in many standard textbooks (e.g., \cite{gregory}) and is being used extensively for the detection of various sources in the MLDC~\cite{Arnaud2}. An MHMC search constructs a sequence of points in parameter space, ${\bf x_i}$. From a point ${\bf x_i}$, a value for the next point in the chain is proposed by drawing from a proposal distribution $q({\bf y} | {\bf x_i})$. A random parameter is then drawn from a uniform distribution $w\in U[0,1]$ and if $w < {\cal H}$ the move is accepted and the chain moves to ${\bf x_{i+1}} = {\bf y}$, otherwise the move is rejected and ${\bf x_{i+1}}= {\bf x_i}$. Here ${\cal H}$ is the Metropolis-Hastings ratio
\begin{equation}
{\cal H}({\bf y}, {\bf x_i}) = \frac{\pi({\bf x_i})p(s|{\bf y})q({\bf x_i}|{\bf y})}{\pi({\bf y})p(s|{\bf x_i})q({\bf y}|{\bf x_i})},
\end{equation}
$\pi({\bf x_i})$ is the prior on the parameters and $p(s|{\bf x})$ is the likelihood of the data $s$ given the parameters ${\bf x}$. For the time-frequency MHMC search, we used priors that were uniform within the parameter ranges specified for the MLDC, and based the log-likelihood on the overlap between the observed whitened time-frequency spectrogram and a theoretical spectrogram built from the AK model (see \cite{CATS} for additional details). Our proposal distribution was constructed from the Fisher Information Matrix for our waveform model.

For the Round 1B analysis, the MHMC chains were seeded at the rough parameters estimated in the first stage and were allowed to run for $10^5$ points. The search used plunge eccentricity and plunge time as parameters rather than initial frequency and eccentricity, with a prior on the plunge time estimated from the detected time-frequency tracks. The MHMC chains refined the initial parameter estimates, but did not move far from the initial parameters, as is clear by comparing the ``Rough'' and ``Final'' parameters in Table~\ref{table:tfparams}. The chains moved somewhat further for 1B.3.5, for which a mis-identification of the harmonics led to a poor initial parameter estimation. We also ran chains that started at random points in the parameter space, and these converged to the same parameter estimates, which gave us confidence that the code was working as expected.

\section{Results}\label{sec:results}
In Table~\ref{table:tfparams} we tabulate the true and recovered parameters for our search of the MLDC Round 1B data. In the table ``Rough'' denotes the first, rough parameter estimates, ``Final'' denotes the refined estimates obtained from the MHMC search and ``True'' denotes the true parameters. We see that, overall, we obtained good estimates of all the parameters, except for the lowest SNR challenge, 1B.3.5. The ``Rough'' parameter estimation was already quite good in most cases, but the parameter estimation was further improved by the MHMC, except in the case of challenge 1B.3.1 where the final parameters were worse than the rough parameters. Our sky position estimator has a degeneracy between antipodeal sky positions $(\beta, \phi_S)$ and $(-\beta, \phi_S+\pi)$, which reflects the real degeneracy between these points in the low frequency approximation~\cite{cutler98}. We used the convention that we returned values with $\beta \in [0,\pi/2]$, which meant that we returned the antipodeal position for sources 1B.3.4 and 1B.3.5. The sky position estimates were otherwise very good. We make some source specific comments below.

\begin{table}[htb]   
\begin{tabular}{|cc|c|c|c|c|c|c|c|c|}\hline
\multicolumn{2}{|c|}{Challenge}& $\beta$&$\phi_S$&S&$m/M_\odot$ &$M/M_\odot$ 
	& $\nu_0$ (mHz) & $e_0$ & $\lambda$\\
\hline
&Rough & 0.491 & 4.9 & 0.698 & 10.5 & 9.56 $\times 10^6$ & 0.1920 & 0.2102 & 0.424\\
1B.3.1&Final & 0.4941 & 4.939 & 0.6667 & 10.404 & 9.787 $\times 10^6$ & 0.1921 & 0.1886 & 0.1938\\
&True & 0.5526 & 4.9104 & 0.6982 & 10.2961 & 9.5180 $\times 10^6$ & 0.1920 & 0.2144 & 0.4395\\
\hline
&Rough & 0.393 & 4.59 & 0.641 & 9.593 & 5.118 $\times 10^6$ & 0.3422 & 0.2251 & 1.48 \\
1B.3.2&Final & 0.4028 & 4.656 & 0.6371 & 9.817 & 5.250 $\times 10^6$ & 0.3423 & 0.2017 & 1.423\\ &True & 0.3597 & 4.6826 & 0.6380 & 9.7711 & 5.2156 $\times 10^6$ & 0.3423 & 0.2079& 1.4358\\
\hline
&Rough & 1.100 & 0.628 & 0.525 & 9.713 & 5.2285 $\times 10^6$ & 0.3426 & 0.1958 & 0.921 \\
1B.3.3&Final & 1.013 & 0.7348 & 0.5318 & 9.756 & 5.250 $\times 10^6$ & 0.3426 & 0.1936 & 0.9091\\
&True & 0.9817 & 0.7097 & 0.5333 & 9.6973 & 5.2197 $\times 10^6$ & 0.3426 & 0.1993 & 0.9282\\
\hline
&Rough & 1.508 & 1.19 & 0.62 & 10.04 & 9.51 $\times 10^5$ & 0.8473 & 0.4598 & 1.663\\
1B.3.4&Final & 0.7805 & 4.168 & 0.6133 & 9.989 & 9.518 $\times 10^5$ & 0.8512 & 0.4564 & 1.658 \\
&True & -0.9802 & 0.9787 & 0.6251 & 10.1047 & 9.558 $\times 10^5$ & 0.8514 & 0.4506 & 1.6707\\
\hline
&Rough & 1.493 & 0.565 & 0.610 & 9.925 & 9.504 $\times 10^5$ & 0.8823 & 0.4435 & 1.39\\
1B.3.5&Final & 0.726 & 4.45 & 0.618 & 10.49 & 9.500 $\times 10^5$ & 0.887 & 0.417 & 1.47\\
&True & -1.1092 & 1.0876 & 0.6583 & 9.7905 & 1.0334 $\times 10^6$ & 0.8322 & 0.4269 & 2.3196\\\hline
\end{tabular}
\caption{Time-frequency search results for challenges 1B.3.1-1B.3.5.}
\label{table:tfparams}
\end{table}

\subsection{Challenge 1B.3.1}
The high SMBH mass meant that there were a number of near-degeneracies in the parameter space for this challenge, which made individual parameters difficult to resolve despite the high SNR. This problem was exacerbated by the fact that the inclination was such that only one bright $k$-sideband was seen for each $n$-harmonic. Although other, weak, sidebands were detected, the MHMC moved towards lower eccentricity and inclination since this put more of the signal power into the brightest harmonic and hence ``improved'' the fit. In a future implementation of the search, we plan to include a penalty associated with parameter space points that do not reproduce all the tracks that have been seen. The final parameters of the MHMC search were such that the strongest detected track for the $n=2$ harmonic should be the $m=2$ sideband, and the secondary track should be the $m=1$ sideband.  However, the values of $\lambda$ and $\kappa$ were such that the $m=-2$ sideband should have been brighter than the $m=1$ sideband.  Since we could not detect another sideband in the presumed location of $m=-2$, we manually changed the inclination angle $\kappa$, which decouples from the other parameters, to get a fit to the detected tracks.  The resulting estimates for $\lambda$ and $\kappa$ were quite poor as a result. This was the only challenge in which the rough parameter estimates were better than the final parameter estimates. In future searches, we may use only the rough parameter estimation for high mass SMBH sources as it is then easier to control degeneracies.

\subsection{Challenge 1B.3.2}
The MHMC search pushed the SMBH mass all the way to the maximum value allowed by the prior, which suggested a lack of accuracy. The initial rough parameter estimate matched the data well, although not quite as well as the final MHMC result.  We submitted both results, although this was artificial in a sense, because we felt the final point was a better fit and only thought it was wrong because we had artificially tight prior constraints on $M$. When the prior was relaxed, the best fit was found to be for a slightly higher central black hole mass, around $5.26\times10^6 M_{\odot}$.

\subsection{Challenge 1B.3.3}
Again, the MHMC search pushed the SMBH mass to the maximum value allowed by the prior.  This was, however, very close to the value detected by the initial rough estimate. Once again, if a less restricted prior was used, the best fit would have been for a slightly higher central black hole mass. For both 1B.3.2 and 1B.3.3, the true black hole mass was quite close to the edge of the prior which is why this issue arose.

\subsection{Challenge 1B.3.4}
Although sections of multiple harmonics and sidebands were detected, the duration of each detected segment was quite short. This suggested that parameter estimation might not be as accurate as usual. However, the final parameters reproduced the detected tracks well, which gave us confidence in the results, and it is clear from Table~\ref{table:tfparams} that parameter recovery was very good.

\subsection{Challenge 1B.3.5}
The tracks were difficult to find because of the low SNR.  Nevertheless, we succeeded in finding a track corresponding to the $n=3$ harmonic with a possible sideband near $t=0$, and tracks corresponding to the $n=2$ and $n=3$ harmonics, each with one sideband, near $t=3.6 \times 10^7$ s.  The short duration of the tracks made accurate parameter estimation difficult, and the best fit parameters did not yield tracks that passed through the tracks detected near $t=0$ (although they were close). We expected these parameters to be the worst of the five Challenges, which is confirmed by Table~\ref{table:tfparams}. However, we were unable to detect anything in the corresponding data set in MLDC Round 2, so the performance of our approach in Round 1B is a significant improvement.

\section{Summary and future plans}\label{sec:future}
The time-frequency analysis of the Round 1B data was very successful and our entry had the most complete parameter recovery out of the three submissions for this challenge. The other challenge entries used MHMC matched filtering. Ultimately that will be more sensitive and have much better parameter recovery than time-frequency analysis, but the time-frequency approach clearly has some parameter estimation power and will potentially be a useful first step in an analysis pipeline. There are a few ways in which the search could be improved. The CATS algorithm, as applied to Round 1B, was not automatic, but interesting regions of parameter space were selected either by eye or by using the output of a HACR search. In the future, we will assess the performance of the CATS algorithm without using prior information. We will also compare the Radon transform algorithm outlined earlier against the CATS algorithm. The parameter extraction cannot be significantly improved. However, the template spectrogram model used in Round 1B to evaluate the likelihood in the MHMC search is only an approximation. It needs to be properly compared against mock data and might be improved. We also want to investigate imposing penalties in the likelihood evaluation to ensure the parameters predict all the tracks that are detected in the spectrogram. It should also be possible to use the total power in the tracks to estimate the total SNR or equivalently the distance to a source, which we have not done so far. The parameter estimation that we have done to date has been very specific to the AK model used for the MLDC data. The general techniques will generalise to more accurate waveform models but the mapping between harmonic frequencies and physical parameters will change. We plan to derive analogous methods for parameter extraction from ``numerical kludge'' waveforms~\cite{GG06,NK} as a first step towards this non-trivial generalisation.

So far, the MLDC EMRI data sets have each contained a single EMRI signal in instrumental noise. The prior that there is only one signal in the data set is a very powerful one, but also very unrealistic. The true LISA data stream will contain a few tens to as many as a thousand EMRI sources~\cite{gairetal} plus signals from white dwarf binaries and supermassive black hole mergers. Time-frequency detection algorithms, plus the simple parameter extraction techniques that we have used to date, will have limited effectiveness in such situations. However, it is a useful exercise to see at what point these techniques no longer work, and it is likely that the methods will still be able to detect the brightest sources even in the presence of confusion, although track detection thresholds will have to be modified to account for the confusion ``noise''. The MLDC Round 3 will begin to address these issues, as there will be only one EMRI data set containing five overlapping sources. We will explore confusion while attempting to analyse Round 3. However, the sources will also be low SNR, so it is not yet clear whether time-frequency techniques will be able to detect anything. If necessary, we will create similar data containing brighter sources. We also plan to explore the efficiency of these methods when applied to data containing a single EMRI with a reduced galaxy to mimic the effect of confusion from white-dwarf binaries.

\acknowledgments
JG thanks the Royal Society for support and the Albert Einstein Institute for hospitality and support while part of this work was being completed. IM was partially supported by NASA ATP Grant NNX07AH22G to Northwestern University. LW's work is supported by the Alexander von Humboldt Foundation's Sofja Kovalevskaja Programme funded by the German Federal Ministry of Education and Research.

\end{document}